\documentclass{procmult}

\usepackage{graphicx}
\usepackage{amsmath}
\usepackage{amsfonts}
\usepackage{hyperref}
\usepackage{amssymb}
\usepackage{txfonts}

\begin{document}

\title{Detecting the Cosmological Stochastic Background of Gravitational Waves with FastICA}
\titlerunning{Detecting CSB-GW with FastICA}
\authorrunning{Izzo L. et al}
\author{Luca Izzo\inst{1,2,3}, Salvatore Capozziello\inst{1} and MariaFelicia De Laurentis\inst{1}}

\institute{Dipartimento di Scienze Fisiche, Universit\`a di Napoli
"Federico II" and INFN Sez. di Napoli, Compl. Univ. Monte S.
Angelo, Ed. N, Via Cinthia, I-80126 Napoli, Italy,  \and ICRANet
and ICRA, Piazzale della Repubblica 10, I-65122 Pescara, Italy,
\and Dip. di Fisica, Universit\`a di Roma "La Sapienza", Piazzale
Aldo Moro 5, I-00185 Roma, Italy}

\maketitle
\abstract {Abstract}
We show that the stochastic background of gravitational waves, produced in the early cosmological epochs, strictly depends on the assumed theory of gravity. In particular, the specific form of the function f(R), where R is the Ricci scalar, is related to the evolution and the production mechanism of gravitational waves. Using a neural network algorithm which only requires non-Gaussian nature and independence of the input signals we conclude that, in order to detect a CSB signal, the interferometric sensitivity of detector like VIRGO will be improved.

\section{Introduction to the Cosmological Stochastic Background of Gravitational Waves (CSB)}

In 1954 Penzias and Wilson, \cite{Penzias}, using an antenna for phone communication, discovered an isotropic signal that today is well-known as the Cosmological Microwave Background Radiation (CMBR). Today it is commonly accepted as the electromagnetic relic signature of the Big Bang: it has the characteristics of a black-body radiation at 3 K. In the same way there would be a relic graviton background originated in the Big Bang and the detection of this gravitational signal should provide more information on the very early stages of the Universe like the Inflation.

It is well known that the inflation is an epoch in which the Universe was characterized by an accelerated expansion, likely due to a phase transition of a scalar field. In particular it was shown by Abbott and Wise, \cite{Abbott}, that the perturbation spectra is the signature of some vacuum fluctuations in the early Universe and these fluctuations could have produced also gravitational radiation, providing a relic signal of Gravitational Waves. The mathematical transposition of this process can be given in terms of $f(R)$-theories, simply using the formalism of the conformal transformations. In this way, following the work of \cite{Capozziello}, it would be possible estimate a similar signal in our epoch. Moreover from the amplitude of the CSB-GW signal it is possible estimate the average amplitude of the primordial fluctuations, so the detection of a relic signal of GW is more than a simple challenge.

\section{Characterization and detection of a CSB of GW signal}    

In order to simulate a possible CSB-GW signal, we must account for several characteristics of this hypothetic signal. This signal in fact should be isotropic, stationary and unpolarized so that its principal property is the frequency spectrum. There are several quantities that describe in a more or less precise way this signal: the spectral density $S_h(f)$, the characteristic amplitude $h_c(F)$ and the energy density per unit of frequency $\Omega_{GW}(f) = \frac{1}{\rho_c}\frac{d \rho_{GW}}{d \log{f}}$, where $f$ is the frequency of the GW signal, \cite{Maggiore}. All of these quantities are related with each other, but in order to understand the effects on a detector we need to think in terms of amplitude so, using the TT-gauge and making a sort of time averaging of the gravitational perturbation detected, $h_{ab}(t)$, we obtain the important results that:
\begin{equation}
 <h_{ab}(t) \, h^{ab}(t)> = 2 \int_{f=0}^{f = \infty} d(\log{f})h_c^2(f).
\end{equation}

We can also obtain the contribute in terms of energy density of the gravitational perturbation. If we make a spatial averaging over several wavelenghts we obtain the density of stochastic GWs $\rho_{GW}$, which relates directly with the characteristic amplitude. Now dividing by the critical density of the Universe, we obtain:
\begin{equation}
 \Omega_{GW}(f) = \frac{2 \pi^2}{3 H_0^2} f^2 h_c^2(f),
\end{equation}
which represents a numerical contribution to the total density parameter of the universe $\Omega$. Note that, as we will see, $h_c$ is very small, so this contribution is negligible in the well-known cosmological consideration about the total density of the Universe.

Actually, $h_c(f)$ is not yet the most useful dimensionless quantity to use for the comparison with experiments. In
fact, any experiment involves some form of binning over the frequency. In a total observation time $T$ , the resolution in frequency is $\Delta f = 1/T$ , so it is convenient to define:
\begin{equation}
 h_c (f, \Delta f) = h_c (f) \left(\frac{\Delta f}{f}\right)^{1/2},
\end{equation}
and using $1/(1yr) \simeq 3.17 \times 10^{-8} Hz$ as a reference value for $h_0 \Omega_{GW}(f)$, we finally have the expression for the characteristic amplitude:
\begin{equation}
h_c(f, \Delta f) \simeq 2.249 \times 10^{-25}\left(\frac{1 Hz}{f}\right)^{3/2} \left(\frac{h_0 \Omega_{GW}(f)}{10^{-6}}\right)^{1/2} \left(\frac{\Delta f}{3.17 \times 10^{-8}Hz}\right)^{1/2}.
\end{equation}

This quantity matches directly with several constraints given by some experiment by satellite as COBE and WMAP, but also from the Big Bang Nucleosynthesis theory, for which we would have
\begin{equation}
 h_c(f) < 2.82 \times 10^{-21} \left(\frac{1 Hz}{f}\right).
\end{equation}
Moreover another constraint could be given by the Sachs-Wolfe effect for which
\begin{equation}
 \Omega_{GW} < \left(\frac{H_0}{f}\right)^2 \left(\frac{\delta T}{T}\right)^2,
\end{equation}
where $\delta T$ corresponds to the anisotropies of the temperature in the CMBR.

Now the detected signal is directly connected with the gravitational perturbation by $s(t) = D^{ab}h_{ab}(t)$, where $D^{ab}$ is the detector tensor which depends on the directions of the GW-interferometer arms. However the output signal will present an additional component due to the interferometer noise so our simulated signal will be $S(t) = n(t) + s(t)$, where $n(t)$ is the noise signal component. Using a pre-compiled MATLAB algorithm, \cite{Cuoco}, we reconstructed $n(t)$ for a GW-interferometer like VIRGO, while for $s(t)$ we considered several oscillated (sinusoidal-like) signal; this because we are interested in the simple detection, not in the rivelation of the wave form, of the CSB signal. 

\section{Signal Analysis}

With this in mind we recognize that the reconstructed signal is a chaotic signal, so from the observation of a dynamical variable we want to find a method to reconstruct the phase space, or the number of variables that compose the starting signal. 

For this reason we used the embedding theorem, \cite{Takens}, which states that, given an N-point time series, $x_1, x_2,...,x_N$, the phase space vectors are reconstructed in order to get a signal matrix as
\begin{equation}
\left(
\begin{array}{cccc}
x_1 & x_2 & \cdots & x_{N-(m-1)\tau}\\
x_{1+\tau} & x_{2+\tau} & \cdots & x_{N-(m-1)\tau + \tau}\\
\cdots & \cdots & \cdots & \cdots\\
x_{1+(m-1)\tau} & x_{2+(m-1)\tau} & \cdots & x_N\\ 
\end{array}
\right),
\end{equation}
where $\tau$, the \emph{time delay}, and $m$, the \emph{minimum embedding dimension}, or the dimension of the reconstructed phase space, are two unknown parameters to determine.

In order to determine $\tau$ we minimized the average mutual information $I_T$:
\begin{equation}
I_T = \sum_{i=1} P(x_i, x_{i+T}) log_2 \left[\frac{P(x_i, x_{i+T})}{P(x_i)P(x_{i+T)}}\right], 
\end{equation}
that represents the average amount of information about $x_{i+T}$ conditioned by observations $x_i$. For $m$ we used a method found by Cao, \cite{Cao}, which, without going into details, allows to determine the minimum embedding dimension. 

The knowledge of $\tau$ and $m$ allows us to use a statistical algorithm, the Independent Component Analysis (ICA), \cite{ICA}, derived from neural networks, in order to going back to the single sources which compose the starting signal. If we observe a signal $\mathbf{y} = \mathbf{A} \mathbf{x}$, with $\mathbf{A}$ mixing matrix and $\mathbf{x}$ the unknown sources, that satisfies the following conditions: a) they must be independent, b) they must be non-Gaussian, then we can construct a combination of the observed signals as:
\begin{equation}
 \mathbf{s =w^T y = w^T A x = z^T y}. 
\end{equation}
Now for the Central Limit Theorem $\mathbf{s}$ becomes less Gaussian when equals any one of the $\mathbf{y}$. In general $\mathbf{s}$ is more Gaussian of any $\mathbf{y}$, so we need an estimator of non-Gaussianity. This is given by the Negentropy, which states for negative statistical entropy, $J(s) \propto [E{G(s)} - E{G(\nu)}]$, where $E{}$ is the expectation value of the quantity in consideration and $\nu$ is a null-mean and unitary-variance Gaussian variable. In this way we require a sort of maximization of the Negentropy and we obtain it using an iteration algorithm, named FastICA. 

The algorithm FastICA consists of 2 steps. The first one is an iteration cycle for the determination of a single unit, i.e. a single vector component of the inverse matrix $\mathbf{w}$. The cycle is the following: 
\begin{itemize}
 \item random choice of the starting vector $\mathbf{w}$
 \item $\mathbf{w = E\{y}$g($\mathbf{w^T y}$)$\mathbf{\} - E\{}$g'($\mathbf{w^T y}$)$\mathbf{\}w}$
 \item $\mathbf{ w = w^T / \parallel w^T \parallel}$
 \item if there isn't a convergence, do repeat the cycle
\end{itemize}
where convergence stands for new and old values of $\mathbf{w}$ that point in the same direction. The second step is the determination of the remaining components, for which we used a two-steps Gram-Schmidt algorithm:
\begin{equation}
 \mathbf{w}_{p+1} = \mathbf{w}_{p+1} - \sum_{j=1}^p \mathbf{w}_{p+1}^T \mathbf{w}_j \mathbf{w}_j,
\end{equation}
and
\begin{equation}
 \mathbf{w}_{p+1} = \mathbf{w}_{p+1} / \sqrt{\mathbf{w}_{p+1}^T \mathbf{w}_{p+1}},
\end{equation}
that therefore allows us to recover $\mathbf{w}$ and therefore the mixing matrix $\mathbf{A}$, from which, and from $\mathbf{y}$, we come back to the single components $\mathbf{x}$ of our starting signal $S(t)$.

\begin{figure}\label{fig:no1}
\centering
\includegraphics[width=11cm, height=7.5 cm]{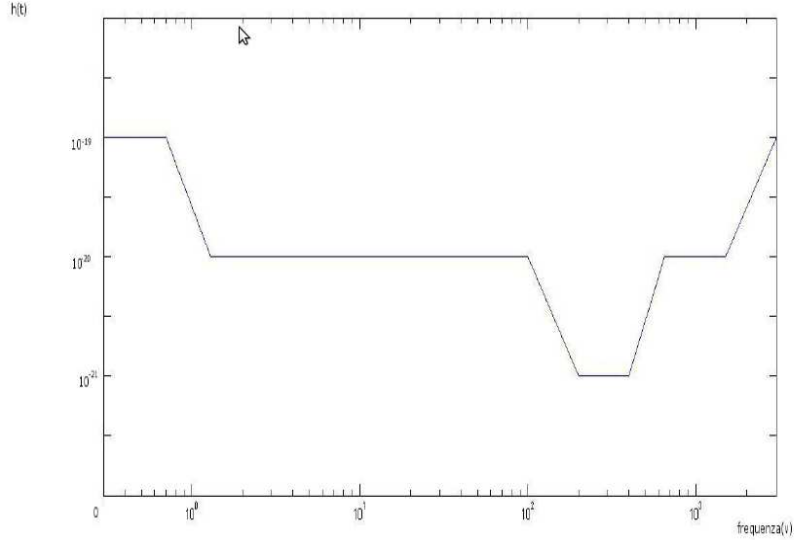}
\caption{Minimum detectable threshold by an interferometer VIRGO-like for an hypothetical CSB signal of GW, with frequency increasing. The maximum sensitivity is obtained for signals of frequency in the range $(100-1000 Hz)$.}
\end{figure}

So we have constructed an algorithm for the estimate of the independent component of an hypothetical CSB-GW VIRGO-detected signal. In this way, using our procedure, we can determine a minimum detectable threshold for the CSB-GW signal, giving to the gravitational signal $s(t)$ an oscillatory behaviour varying with the frequency. The results of our analysis are shown in figure \ref{fig:no1} and in the table \ref{table:no1}. We immediately see that the maximum sensitivity is obtained for signals of frequency in the range $(100-1000 Hz)$, but more important we note that our algorithm works very well with this type of signals also if the identification of the estimated signals were done with a graphical recognition.

\begin{table}
\caption{Results of the simulations} 
\label{table:no1} 
\centering 
\begin{tabular}{c c} 
\hline\hline 
Frequency ($\nu$) & Amplitude $h(t)$ \\ 
\hline 
 $0.3$ & $10^{-19}$ \\
 $0.7$ & $10^{-19}$ \\ 
 $1.3$ & $5 \times 10^{-20}$ \\
 $2.5$ & $5 \times 10^{-20}$ \\
 $6.5$ & $5 \times 10^{-20}$ \\
 $100$ & $5 \times 10^{-20}$ \\
 $200$ & $10^{-20} \sim 5 \times 10^{-20}$ \\
 $400$ & $10^{-20}$ \\
 $650$ & $5 \times 10^{-20}$ \\
 $1500$ & $5 \times 10^{-20} \sim 10^{-19}$ \\
 $3000$ & $10^{-19}$ \\
\hline 
\end{tabular}
\end{table}

\section{Conclusions}

From our analysis we conclude that it is necessary to constraining the theory for a Cosmological Stochastic Background of Gravitational Waves and we advise to deepen the relations among the Sachs-Wolfe effect and the gravitational perturbations in $f(R)$ theories. This could be a starting point for the theoretical analysis of the detection with interferometer of a gravitational signal coming from the relic background originated from the Big Bang. 

Moreover also the application to VIRGO-detected signals of the ICA methods needs to be improved, in particular using a time series longer, a number of the sources more than 2 and possibly the utilization of a “less-human” method for the identification of gravitational signals. The last, but not the least, we suggest a frequency analysis in
order to use directly the energy density for the study of the connections with the Sachs-Wolfe effect.

\end{document}